\documentclass[reprint,
amsmath,amssymb,
aps,
]{revtex4-2}
\usepackage{graphicx}
\usepackage{dcolumn}

\begin{document}

\title{
Extraordinary $\pi$-Electron Superconductivity\\
Emerging from a Quantum Spin Liquid
}

\author{
S. Imajo$^{1,2,*}$,
S. Sugiura$^{3,4}$,
H. Akutsu$^1$,
Y. Kohama$^2$,
T. Isono$^{3}$,
T. Terashima$^3$,
K. Kindo$^2$,
S. Uji$^3$,
and
Y. Nakazawa$^1$
}

\affiliation{
$^1$Graduate School of Science, Osaka University, Toyonaka, Osaka 560-0043, Japan\\
$^2$Institute for Solid State Physics, University of Tokyo, Kashiwa, Chiba 277-8581, Japan\\
$^3$National Institute for Materials Science, Tsukuba, Ibaraki 305-0003, Japan\\
$^4$Present address: Institute for Materials Research, Tohoku University, Sendai 980-8577, Japan\\
}
\date{\today}

\begin{abstract}
Quantum spin liquids (QSLs), in which spins are highly entangled, have been considered a groundwork for generating exotic superconductivity.
Despite numerous efforts, superconductivity emerging from QSLs has been unrealized in actual materials due to the difficulties in stabilizing QSL states with metallic conductivity.
Recently, an organic compound, $\kappa$-(BEDT-TTF)$_4$Hg$_{2.89}$Br$_8$, with a nearly regular triangular lattice of molecular dimers was recognized as a candidate for doped QSLs.
In this study, we report an unusual superconducting phase of $\kappa$-(BEDT-TTF)$_4$Hg$_{2.89}$Br$_8$: unexpectedly large ratios of the upper critical field to the critical temperature $H_{\rm c2}$/$T_{\rm c}$ in fields not only parallel but also perpendicular to the two-dimensional conducting layers and a very wide region of fluctuating superconductivity above $T_{\rm c}$.
Our results reveal that these peculiarities arise from strong electron correlations and possible quantum criticality unique to the doped QSL state, leading to a heavy mass of itinerant carriers and a large superconducting energy gap.
\end{abstract}

\maketitle
Carrier doping into quantum spin liquids (QSLs) is expected as a possible pathway to realize exotic superconductivity\cite{1}.
In QSLs, novel superconductivity is potentially induced by unconventional mechanisms, where geometrical spin frustration in addition to antiferromagnetic (AF) spin fluctuations play essential roles.
In the early stage of the arguments for high-$T_{\rm c}$ cuprates, AF spin fluctuation in the resonating valence bond (RVB) state was proposed as a possible pairing scenario\cite{1,2,3}, although the cuprates possess no geometrical frustration.
After that, other unconventional pairing mechanisms different from the RVB model were discussed in the framework of Hubbard physics\cite{4}.
The situation resembles that for heavy fermions\cite{5} and organics\cite{6}, and therefore, the origin of the unconventional superconductivity is described by the AF fluctuations in a nonfrustrated lattice, even though it may be related to other degrees of freedom.
Despite various theoretical predictions to reveal the effect of geometrical spin frustrations on superconductivity\cite{2,7,8}, no detailed experimental results indicative of the frustration effect on superconductivity have been reported until now.

\begin{figure}
\begin{center}
\includegraphics[width=\hsize,clip]{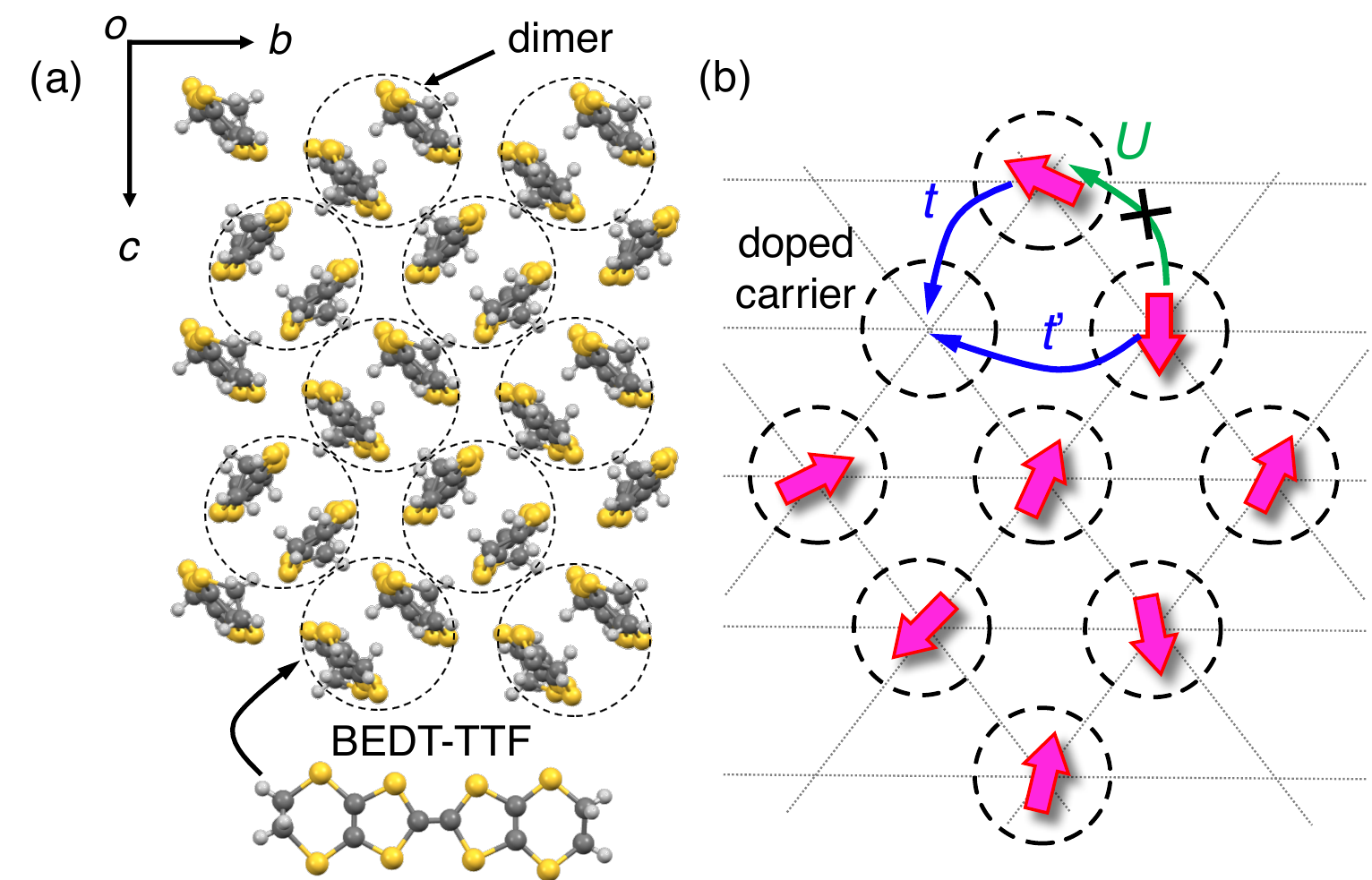}
\end{center}
\caption{(Color online) Crystal structure of $\kappa$-HgBr.
(a) Arrangement of BEDT-TTF molecules in $\kappa$-HgBr in the $b$-$c$ plane.
The broken circles represent the BEDT-TTF molecule dimers.
(b) Schematic of (a).
The pink arrows show the localized spins.
The blue arrows express the interdimer transfer integrals, $t$ and $t^{\prime}$.
The green arrow depicts the on-site Coulomb repulsion $U$ prohibiting double occupancy of electrons at one site.
}
\label{fig1}
\end{figure}

Recently, Oike et al.\cite{9} reported that one of the $\kappa$-type organic superconductors, $\kappa$-(BEDT-TTF)$_4$Hg$_{2.89}$Br$_8$ (abbreviated $\kappa$-HgBr hereafter), is a promising candidate for doped QSLs.
As depicted in Fig.~\ref{fig1}(a), the BEDT-TTF dimers, forming $\pi$-electron conducting layers, are arranged in a triangular lattice characterized by $t^{\prime}$/$t$$\sim$1, where $t$ and $t^{\prime}$ represent the transverse and diagonal transfer integrals in Fig.~\ref{fig1}b.
Band-structure calculations based on the extended H${\rm \ddot{u}}$ckel method and tight-binding approximations with the dimer model\cite{9,10,11} suggest that the electron correlation $U$/$t$, where $U$ represents the on-site Coulomb repulsion, is 1.5-2 times larger than those for other well-known $\kappa$-type salts, such as $\kappa$-(BEDT-TTF)$_2$Cu$_2$(CN)$_3$ ($\kappa$-Cu$_2$(CN)$_3$) and $\kappa$-(BEDT-TTF)$_2$Cu[N(CN)$_2$]Br ($\kappa$-Br).
The large $U$/$t$ suggests that $\kappa$-HgBr should be a Mott insulator.
Indeed, the magnetic susceptibility of $\kappa$-HgBr has a broad maximum, which is well described by a triangular lattice QSL with an AF interaction.
The behavior is very similar to that of other dimer-Mott insulating QSLs, such as $\kappa$-Cu$_2$(CN)$_3$\cite{12}, $\beta$$^{\prime}$-EtMe$_3$Sb[Pd(dmit)$_2$]$_2$\cite{13} and $\kappa$-H$_3$(Cat-EDT-TTF)$_2$\cite{14}.
However, the electrical transport in $\kappa$-HgBr exhibits a metallic nature and superconductivity below $\sim$4~K\cite{9,10,11,15,16}, which is significantly distinct from the insulating behavior of the other QSLs.
This phenomenon occurs because the incommensurate mercury HgBr chains, forming insulating layers, dope nonstoichiometric hole carriers into the conducting layers.
Raman spectroscopy\cite{17} detects a carrier-doped state with a fractional charge of the molecule (BEDT-TTF)$_2^{+1.11}$ according to the anion charge Hg$_{2.89}$Br$_8^{-2.22}$.
In addition to the anomalous decoupling of the magnetism and conductivity, Oike et al. reported a possible pressure-induced quantum phase transition leading to non-Fermi liquid behavior\cite{9,11}.
Although the normal state was well examined in these previous studies, the details of the superconductivity emerging from the doped QSL are still unclear.
Therefore, we performed various measurements in high fields to understand the relation between the frustrated QSL state and the superconductivity.

\begin{figure}
\begin{center}
\includegraphics[width=\hsize,clip]{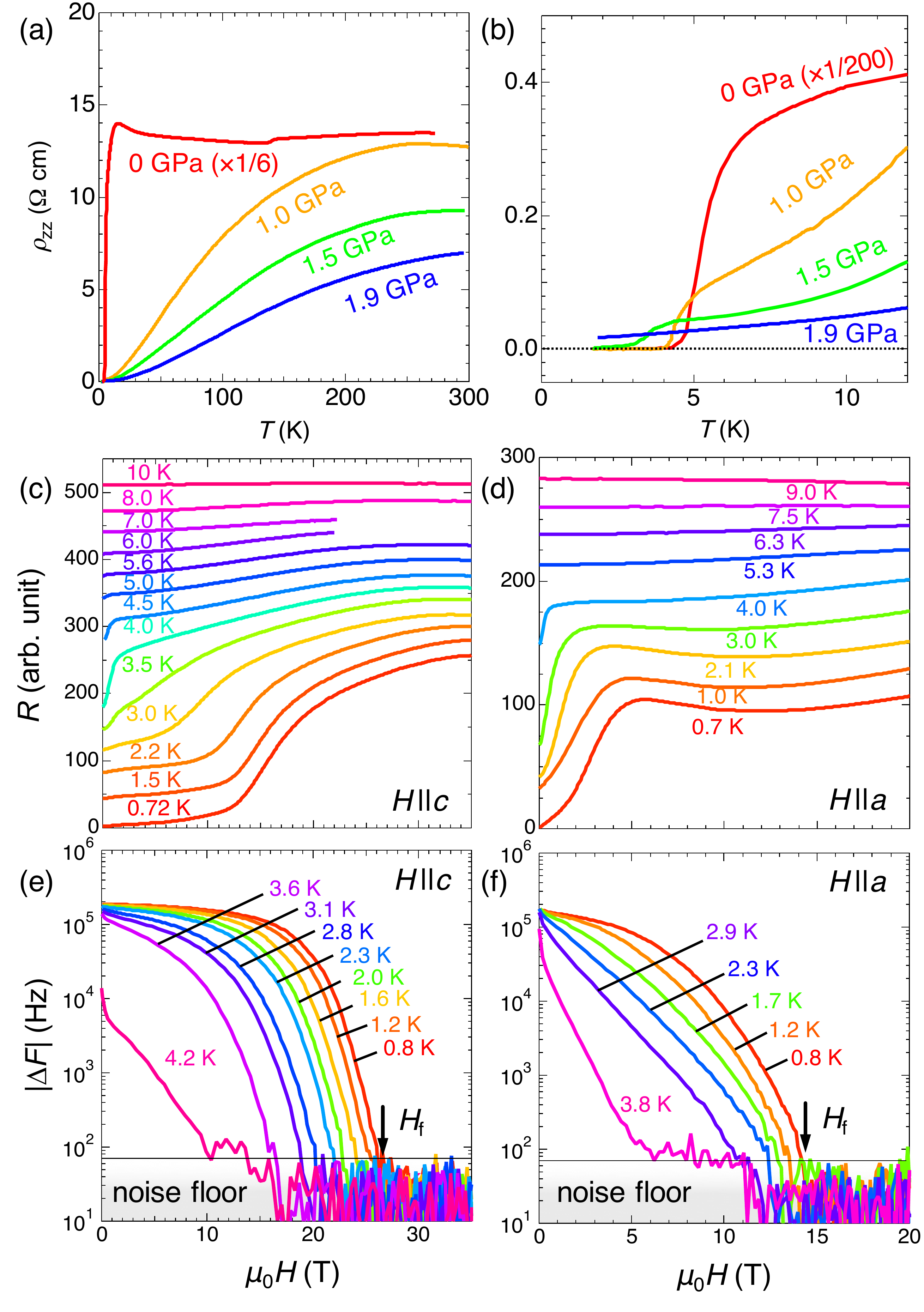}
\end{center}
\caption{(Color online) Transport properties of $\kappa$-HgBr.
(a),(b) Temperature dependences of the out-of-plane resistance under each pressure.
An enlarged plot of the low temperature region is shown in (b).
(c),(d) Out-of-plane field-dependent resistance of $\kappa$-HgBr at each temperature in fields parallel (c) and perpendicular (d) to the conducting plane.
These data include some offsets for clarity.
(e),(f) Magnetic field dependences of the change in the resonant frequency of the radiofrequency TDO measurement in the parallel fields (e) and perpendicular fields (f).
}
\label{fig2}
\end{figure}

 Figures~\ref{fig2}(a) and \ref{fig2}(b) show the temperature dependences of the out-of-plane resistivity at ambient pressure, 1.0, 1.5, and 1.9~GPa.
The resistivity at ambient pressure (0~GPa) is almost constant down to 20~K and then shows a maximum at $\sim$15~K with a small upturn.
As shown in Fig.~\ref{fig2}(b), the ambient pressure resistivity gradually decreases below $\sim$7~K and reaches zero resistivity at $T_{\rm c}$$\sim$4.3~K due to the superconducting transition.
The non-Fermi liquid behavior is suppressed at high pressures, and the conventional Fermi liquid behavior $\rho$$_{zz}$$\propto$$T^2$ appears, as reported in earlier works\cite{9,10,11}.
The small residual resistivity value $\rho$$_{zz}$(0~K) and large residual resistivity ratio $\rho$$_{zz}$(300~K)/$\rho$$_{zz}$(0~K) of the Fermi liquid state (e.g., $\sim$400 at 1.9~GPa) verify the high sample quality, and therefore, the non-Fermi liquid nature at ambient pressure comes from intrinsic origins, not from impurities or defects.
The magnetic field dependence of the resistance at ambient pressure is displayed in Figs.~\ref{fig2}(c) and \ref{fig2}(d).
The magnetic field direction along the $a$-axis, $H$$\parallel$$a$, is perpendicular to the two-dimensional conducting plane, while $H$$\parallel$$b$ and $H$$\parallel$$c$ are in-plane magnetic fields, as presented in Fig.~\ref{fig1}(a).
Despite the high sample quality, the resistive transition of the superconductivity is rather broad, leading to inherent ambiguity in the determination of the upper critical field $H_{\rm c2}$.
However, in the field derivative curve d$\rho$/d$H$, we observe a characteristic deviation from the smooth field dependence below $H_{\rm f}$, which can likely be defined as the upper boundary of the fluctuating superconductivity (FSC) region\cite{SM}.
We complementarily performed radiofrequency tunnel diode oscillator (TDO) measurements and observed the resonant frequency change $|$$\Delta$$F$$|$ as a function of field, as presented in Figs.~\ref{fig2}(e) and \ref{fig2}(f)\cite{SM}.
Here, $\Delta$$F$ is predominantly governed by the superconductivity because of the low conductivity of the normal state.
Since this technique is very sensitive to the superconducting component, defining $H_{\rm f}$ as the value at which $|$$\Delta$$F$$|$ exceeds the noise floor is reasonable, as indicated by the black arrows.
At 4.2~K for $H$$\parallel$$c$ (Fig.~\ref{fig2}(e)), $|$$\Delta$$F$$|$ remains small even at 0~T, which is suggestive of FSC in almost the whole field region up to $H_{\rm f}$.

\begin{figure}
\begin{center}
\includegraphics[width=\hsize,clip]{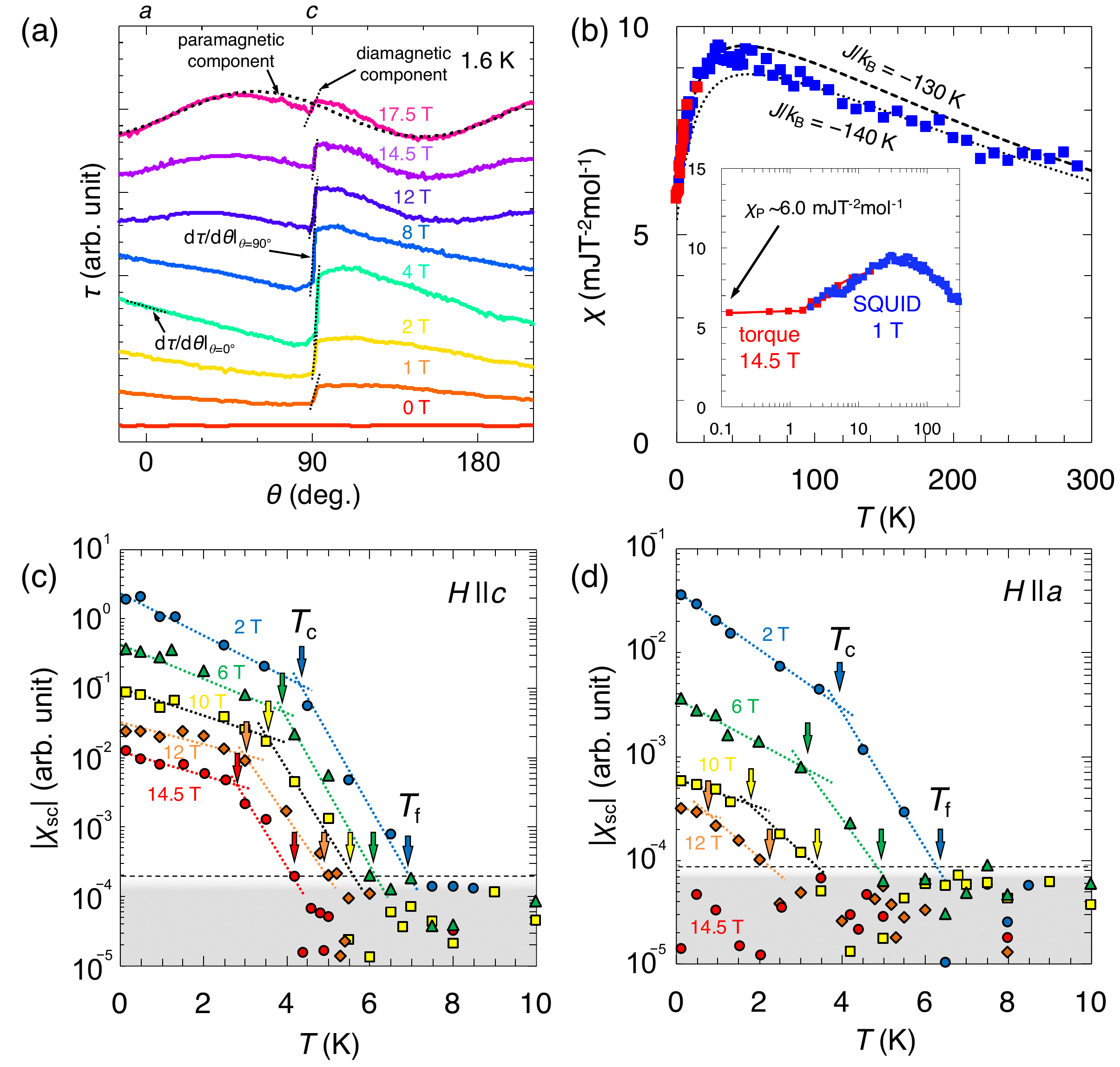}
\end{center}
\caption{(Color online) Magnetic properties of $\kappa$-HgBr.
(a) Magnetic field angle dependence of the torque signal $\tau$($\theta$) at 1.6 K with field rotation in the $a$-$c$ plane.
The dashed curve represents the 180-degree periodicity sin2($\theta$$-$$\theta$$_0$) expected for the paramagnetic component.
The dotted lines show the slope of the diamagnetic component of the superconductivity.
(b) Temperature dependence of the magnetic susceptibility.
The blue and red boxes represent the data measured by SQUID under 1~T and torque measurements under 14.5~T, respectively.
The dashed and dotted lines are the curves calculated from the S=1/2 AF Heisenberg model for an isotropic triangular lattice\cite{18} with spin exchange interactions $J$/$k_{\rm B}$=$-$130~K and $-$140~K, respectively.
The inset shows a semilogarithmic plot of the data.
(c),(d) Temperature dependences of the superconducting diamagnetic susceptibility determined by the slope of the torque signal in fields parallel to the $c$-axis (c) and $a$-axis (d).
}
\label{fig3}
\end{figure}

To evaluate the magnetic properties, we carried out magnetic torque measurements.
The angle dependence of the torque is shown in Fig.~\ref{fig3}(a).
The magnetic torque comprises two components: the paramagnetism originating from the twofold anisotropy of the $g$-value and the diamagnetism due to the superconductivity exhibiting a sharp torque change near 90 deg\cite{SM}.
Figure~\ref{fig3}(b) shows the temperature dependence of the paramagnetic susceptibility obtained from the torque measurements in a perpendicular field of 14.5~T and by a conventional SQUID magnetometer at 1~T.
The susceptibility has a broad maximum at $\sim$30~K, which is well explained by a triangular lattice model\cite{18} with $J$/$k_{\rm B}$=$-$130-$-$140~K, as reported\cite{9}.
We observe constant paramagnetism in the low temperature region down to 0.13~K (inset of Fig.~\ref{fig1}(b)), which is also a typical feature of organic QSLs\cite{12,13,14}.
The paramagnetic susceptibility of the normal state at 0~K $\chi$$_{\rm P}$ is estimated to be $\sim$6.0~mJT$^{-2}$mol$^{-1}$, which is much larger than those of typical organic superconductors (2-4 mJT$^{-2}$mol$^{-1}$; Ref.\cite{6,19}).
The superconducting diamagnetic susceptibility $\chi$$_{\rm sc}$ can be obtained from the slope of the magnetic torque curve\cite{SM}.
The temperature dependences of the absolute value $|$$\chi$$_{\rm sc}$$|$ in parallel and perpendicular fields are shown in Figs.~\ref{fig3}(c) and \ref{fig3}(d), respectively.
As the temperature increases, $|$$\chi$$_{\rm sc}$$|$ decreases, followed by a kink at $T_{\rm c}$, and then steeply decreases down to the noise floor (shaded area) at $T_{\rm f}$, both of which are denoted by arrows.
According to previous torque measurements of organic superconductors\cite{20,21}, we can define $T_{\rm c}$ as the critical temperature at which the superconductivity acquires macroscopic phase coherence, and consequently, the FSC region is between $T_{\rm c}$ and $T_{\rm f}$.
\begin{figure}
\begin{center}
\includegraphics[width=\hsize,clip]{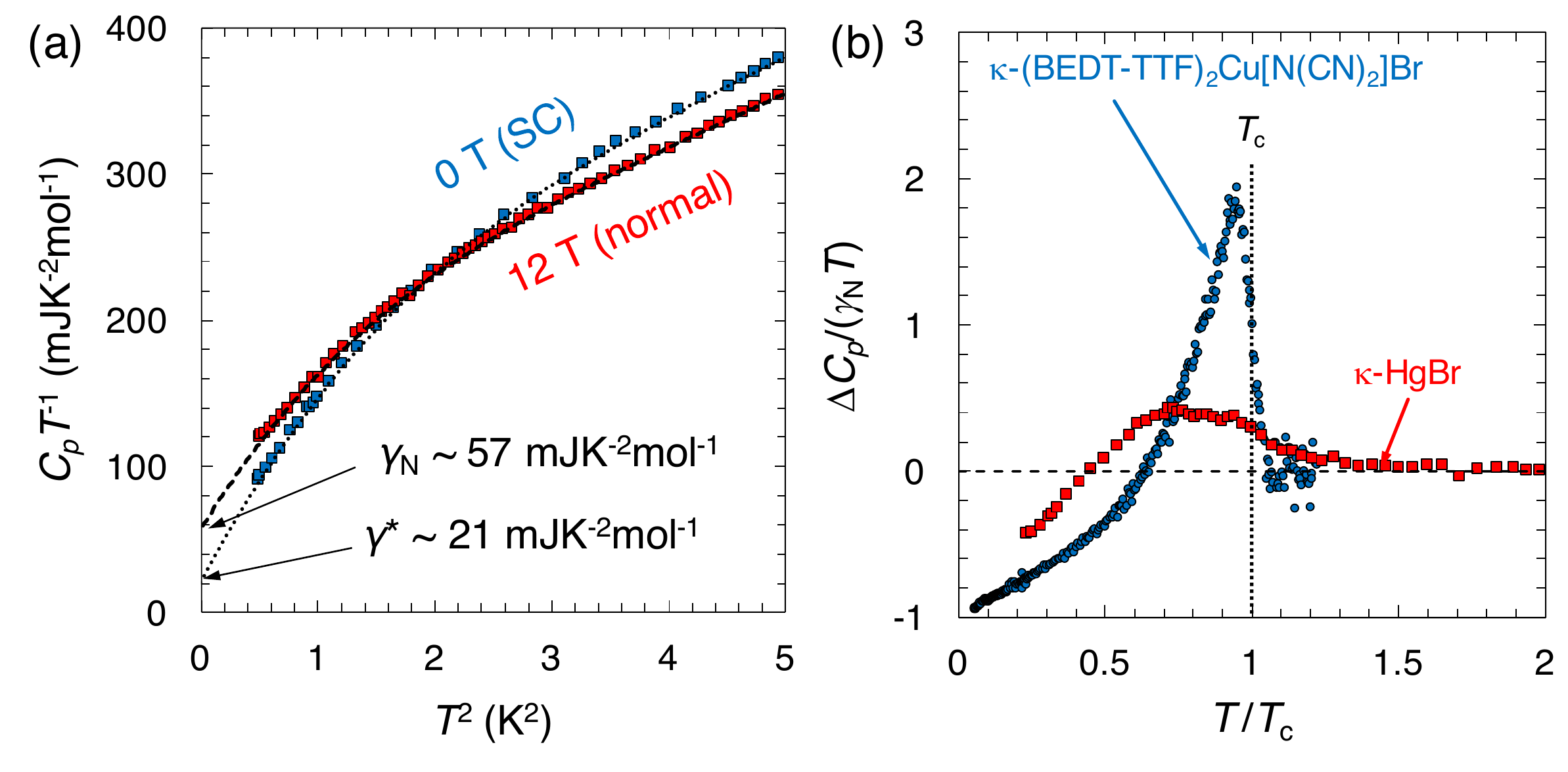}
\end{center}
\caption{(Color online) Thermodynamic properties of $\kappa$-HgBr.
(a) Low-temperature heat capacities at 0~T (blue boxes) and 12~T (red boxes) plotted as $C_pT^{-1}$ vs $T^2$.
The intercepts of the extrapolation represent the electronic heat capacity coefficient $\gamma$$^{\ast}$ at 0~T and $\gamma$$_{\rm N}$ at 12~T.
(b) Reduced heat capacity differences between the 0~T and 12~T data of $\kappa$-HgBr (red) as a function of reduced temperature along with the data of $\kappa$-Br\cite{28} (blue).
The $T_{\rm c}$ used herein is slightly different from that determined by the other measurements due to the difference in the definition\cite{SM}.
}
\label{fig4}
\end{figure}

Figure~\ref{fig4}(a) presents the heat capacity data, $C_pT^{-1}$ vs $T^2$ plot, at 0~T and 12~T in perpendicular fields.
The nonlinear temperature dependence in the plot can likely be ascribed to the low-dimensional phonon contribution of the Hg chains\cite{22}.
The electronic heat capacity coefficient of the normal state $\gamma$$_{\rm N}$ is estimated as $\sim$57~mJK$^{-2}$mol$^{-1}$ by fitting the data above $T_{\rm c}$$\sim$0.8~K at 12~T, which is almost the same as the reported value\cite{22}.
This $\gamma$$_{\rm N}$ value is much larger than those for the dimer-Mott organic compounds (10-30~mJK$^{-2}$mol$^{-1}$; Ref.\cite{23,24}).
The electronic heat capacity coefficient at 0~T, $\gamma$$^{\ast}$, reaches $\sim$21~mJK$^{-2}$mol$^{-1}$, which also far exceeds those observed for other organic superconductors\cite{23,24}.
This large $\gamma$$^{\ast}$ should be of an intrinsic nature because of the high sample quality; however, the detailed origin is currently unclear.
In Fig.~\ref{fig4}(b), the difference between the 0~T and 12~T data $\Delta$$C_pT^{-1}$, corresponding to the electronic heat capacity in the superconducting state, is shown.
For comparison, we also present the data of $\kappa$-Br, which are well explained by a simple extension of the BCS theory, the $\alpha$-model\cite{25,26}.
The broad superconducting transition for $\kappa$-HgBr, which is remarkably different from that for conventional BCS superconductors, is no longer described by the mean-field theory and is consistent with the broad resistive transition.

\begin{figure}
\begin{center}
\includegraphics[width=\hsize,clip]{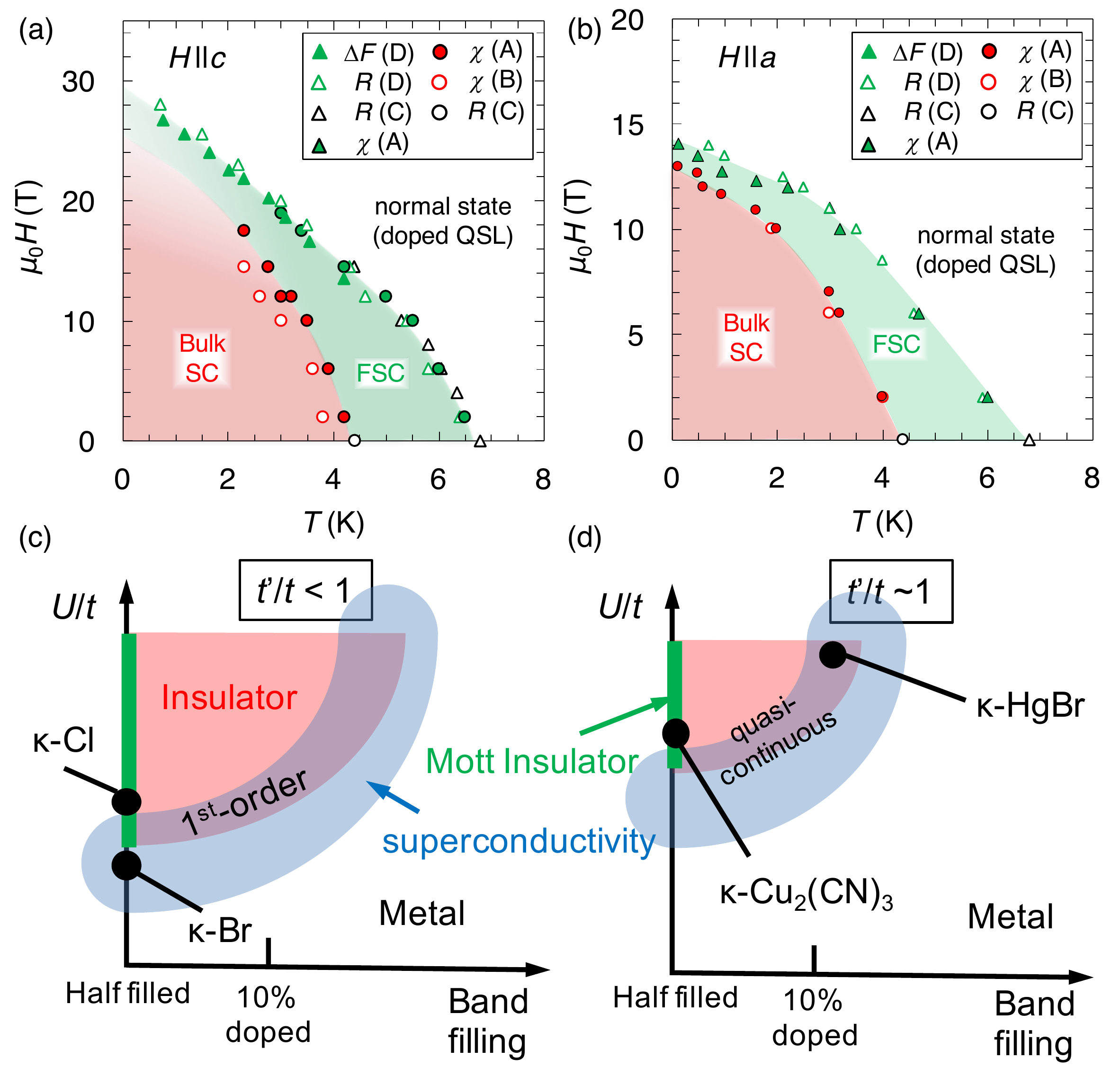}
\end{center}
\caption{(Color online) Superconducting phase diagram of $\kappa$-HgBr and electronic phase diagram of $\kappa$-type salts. 
(a),(b) $H$-$T$ superconducting phase diagrams in parallel (a) and perpendicular (b) fields.
The green and red shaded areas represent the FSC and bulk superconductivity (Bulk SC) regions, respectively.
The circles and triangles denote the $T_{\rm c}$ and $T_{\rm f}$ at each field ($H_{\rm c2}$ and $H_{\rm f}$ at each temperature).
The $T_{\rm c}$ and $T_{\rm f}$ at 0~T are defined by the temperatures at zero resistivity and the onset of the transition, respectively.
The $T_{\rm c}$ in fields is determined by the kink in the temperature dependence of the diamagnetism detected based on the magnetic torque, whereas $T_{\rm f}$ is obtained from the onset of the superconducting transition measured by each technique.
The filled, unfilled, and black-edged symbols correspond to different measurement techniques ($\Delta$$F$, $R$, $\chi$) and samples (A, B, C, D), as noted in the legends.
(c),(d) Schematic of the electronic phase diagrams for the (c) distorted square ($t^{\prime}$/$t$$<$1) lattice and (d) triangular lattice ($t^{\prime}$/$t$$\sim$1) $\kappa$-type salts.
These are drawn on the basis of Ref.\cite{11,32,33}.
The green, red, and blue areas denote the Mott insulator, doped insulator, and superconductivity regions.
Each material (black dot) is positioned according to its electron correlation $U$/$t$ and band filling.
$\kappa$-Cl is the abbreviation of $\kappa$-(BEDT-TTF)$_2$Cu[N(CN)$_2$]Cl.
}
\label{fig5}
\end{figure}

Figure~\ref{fig5} shows the superconducting $H$-$T$ phase diagram obtained from the results of the above measurements.
The bulk superconductivity region, characterized by macroscopic phase coherence, is surrounded by the FSC region, where the superconducting order parameter fluctuates.
By extrapolating the data down to 0~K, we obtain the large critical fields $H_{\rm c2}$$\sim$25~T in parallel fields and $\sim$13~T in perpendicular fields despite $T_{\rm c}$=4~K.
The $H_{\rm c2}$ in parallel fields is consistent with the reported phase diagram above 2~K\cite{16}.
Moreover, the ratio of the FSC to bulk SC regions $T_{\rm f}$/$T_{\rm c}$ reaches approximately 1.6 at 0~T, much larger than the value of $T_{\rm c}$/$T_{\rm f}$$\sim$1.2 for the other nondoped less-frustrated organic superconductors $\kappa$-(BEDT-TTF)$_2$Cu(NCS)$_2$ and $\kappa$-Br\cite{20,21}.
Additionally, we observe the FSC region above $H_{\rm c2}$ even near 0~K, showing that quantum fluctuation, in addition to thermal fluctuation, also plays an important role.

As mentioned above, the remarkable features in $\kappa$-HgBr are the robustness of the superconductivity against magnetic fields and the wide superconducting fluctuation region.
At 0~K, the ratio $H_{\rm c2}$/$T_{\rm c}$ reaches approximately 6~TK$^{-1}$ in parallel fields and 3~TK$^{-1}$ in perpendicular fields, far beyond those for the other organic superconductors\cite{27,28}.
The large $T_{\rm f}$/$T_{\rm c}$ value indicates a significant superconducting fluctuation effect, consistent with the broad transitions in the resistance (Fig.~\ref{fig2}) and heat capacity (Fig.~\ref{fig4}(b)).
The magnetic susceptibility is indicative of the QSL state with the triangular lattice despite the conducting nature.
Additionally, in the normal state, large $\chi$$_{\rm P}$ and $\gamma$$_{\rm N}$ values are observed at 0~K.
To discuss $H_{\rm c2}$, two pair-breaking effects, the orbital effect\cite{29} related to the loss of condensation energy via vortex formation and the paramagnetic effect\cite{30} due to Zeeman energy loss, should be taken into account.
In the case of numerous organic superconductors, the orbital effect is highly anisotropic because of the two-dimensionality, while the paramagnetic effect is almost isotropic due to the weak spin-orbital coupling (almost isotropic $g$-factor).
We first discuss $H_{\rm c2}$ in parallel fields, where the orbital effect is negligible and only the paramagnetic effect operates.
For weak-coupling BCS-type superconductors, the ratio of $H_{\rm c2}$ to $T_{\rm c}$ is given by $H_{\rm c2}$/$T_{\rm c}$=$\Delta$$_0$/($\sqrt{2}$$\mu$$_{\rm B}$$T_{\rm c}$)$\sim$1.84~TK$^{-1}$\cite{30}, where $\Delta$$_0$ is the superconducting energy gap at 0~K.
However, $\kappa$-HgBr exhibits a striking enhancement of the ratio, $H_{\rm c2}$/$T_{\rm c}$$\sim$6~TK$^{-1}$ at 0~K.
This indicates the opening of an unexpectedly large energy gap $\Delta$$_0$.
Surprisingly, this value is much larger than $\sim$2.8~TK$^{-1}$, which was the largest value in organic superconductors\cite{28}.
Furthermore, it is comparable to that of the extremely strong-coupling superconductivity observed in artificial two-dimensional Kondo lattices\cite{31}, ultimately discussed as strongly correlated systems.
Even in perpendicular fields, where the orbital effect must be taken into account in addition to the paramagnetic effect, we also obtain a large ratio $H_{\rm c2}$/$T_{\rm c}$$\sim$3~TK$^{-1}$ at 0~K (Fig.~\ref{fig5}(b)).
Since the orbital critical field is given by $H_{\rm c2}$$\propto$($m^{\ast}$$\Delta$$_0$)$^{2}$, large effective mass $m^{\ast}$ and/or $\Delta$$_0$ are/is required for the large $H_{\rm c2}$.
In parallel fields, the large $\Delta$$_0$ is confirmed.
The observed large $\chi$$_{\rm P}$ and $\gamma$$_{\rm N}$ are direct evidence of the strong mass enhancement because of the relations $\chi$$_{\rm P}$, $\gamma$$_{\rm N}$$\propto$$D$($E_{\rm F}$)$\propto$$m^{\ast}$, where $D$($E_{\rm F}$) is the density of states at the Fermi level $E_{\rm F}$.
In this way, the robust superconductivity in magnetic fields is ascribed to the large $m^{\ast}$ and $\Delta$$_0$, indicative of the strong Cooper pairing of the heavy carriers.
This situation is similar to the cases of heavy fermion superconductors, although typical $\pi$-electron systems have light masses.

Why does $\kappa$-HgBr possess extraordinary features, such as a heavy electron mass, a large superconducting energy gap, and strong fluctuation, distinct from those of typical organic superconductors?
Considering the uniqueness of $\kappa$-HgBr, carrier doping and geometrical frustration should be clues.
The electronic phases of half-filled organic conductors can be controlled by band filling and the electron correlation $U$/$t$.
For typical less-frustrated $\kappa$-type conductors with $t^{\prime}$/$t$$<$1, the phase diagram is depicted in Fig.~\ref{fig5}(c)\cite{6,32}, where superconducting $\kappa$-Br is located below Mott insulating $\kappa$-Cl on the half-filled line.
The Mott phase transition is of the 1st order at low temperatures\cite{6,32}, where quantum fluctuations do not develop.
For the QSL salts with $t^{\prime}$/$t$$\sim$1, the Mott phase boundary shifts toward higher $U$/$t$ (Fig.~\ref{fig5}(d)) since the metallic state is more stabilized as the frustration increases\cite{33}.
The earlier report of a half-filled QSL with $t^{\prime}$/$t$$\sim$1, $\kappa$-Cu$_2$(CN)$_3$\cite{34}, suggests that the Mott transition for the QSL state is quasi-continuous at 0~K and that the electron mass is enhanced near the phase boundary due to quantum criticality, which is possibly related to spinon Fermi surface formation\cite{34}.
The similarity of the QSL behaviors of $\kappa$-Cu$_2$(CN)$_3$ and $\kappa$-HgBr\cite{9} and the position of $\kappa$-HgBr in Fig.~\ref{fig5}(d) imply that the non-Fermi liquid-Fermi liquid crossover of $\kappa$-HgBr\cite{9,11} also originates from the same quantum criticality peculiar to the QSL-metal transition.
Therefore, the non-Fermi liquid behavior of $\kappa$-HgBr at ambient pressure indicates the presence of strong quantum fluctuations coming from the quantum criticality, leading to augmentation of the electron mass.
Additionally, we need to note the effect of the electron correlation.
As reported in previous works\cite{9,10,11}, the $U$/$t$ of $\kappa$-HgBr is 1.5-2 times that of typical $\kappa$-type salts.
The strong electron correlation should also contribute to mass renormalization.
In the case of $\kappa$-type superconductors, the magnitude of $\Delta$$_0$ is enhanced with increasing $U$/$t$\cite{6,24,26,28}.
Thus, that the large $U$/$t$ in $\kappa$-HgBr also contributes to the observed large $\Delta$$_0$ is reasonable.
Namely, the anomalously large $H_{\rm c2}$ arises from the cooperation of the strong electron correlation and quantum criticality unique to the QSL state.
For the strong fluctuation effect, we here discuss the macroscopic phase coherence of the superconductivity.
In superconducting states, a conjugate relation between the uncertainty of the Cooper pair number $\Delta$$n$ and phase uncertainty $\Delta$$\phi$, $\Delta$$n$$\Delta$$\phi$$\approx$1, holds.
Since $\Delta$$n$ is inversely related to the Coulomb energy\cite{20,21,35,36}, the phase uncertainty $\Delta$$\phi$ is enhanced by strong electron correlations.
This means that the discussed large $U$/$t$ enhances the phase fluctuation $\Delta$$\phi$ of the superconductivity and thus results in the wide fluctuation region, as seen in the phase diagrams (Fig.~\ref{fig5}).
This picture agrees with the smaller FSC region (the sharper transition) at high pressures (low $U$/$t$)\cite{9,10,11}.
Nevertheless, the extremely wide FSC region still seems remarkable for $\kappa$-HgBr.
Here, we propose a possible effect of the geometrical frustration on the superconducting fluctuation.
As is evident in the QSL-like magnetic behavior, the spins in $\kappa$-HgBr are affected by the triangular frustration.
Since the superconductivity, presumably singlet pairing, is mediated by the AF spin fluctuation, the frustration could disturb the long-range phase coherence of the singlet pairing superconductivity.
Since the significance of the frustration for the fluctuation effect is still unclear, it should be investigated as an issue in future works on superconductivity emerging from QSL states.

 In summary, we find that the superconductivity in the doped QSL $\kappa$-(BEDT-TTF)$_4$Hg$_{2.89}$Br$8$ has extremely high $H_{\rm c2}$ not only for parallel but also for perpendicular field directions and an extremely wide FSC region above $T_{\rm c}$, in marked contrast to the typical half-filled less-frustrated organic superconductors.
 The extraordinary enhancement of $H_{\rm c2}$ originates from the drastic electron mass renormalization and the opening of a large energy gap.
 The strong electron correlation and the quantum criticality unique to the QSL-Fermi liquid transition are elucidated to be the origin of the unique features.
 These findings will provide new insights into the relation between the superconductivity and the strongly correlated doped QSL states.

We thank H. Taniguchi (Saitama University) for giving us the useful information of the crystal growth.

\renewcommand{\thefigure}{S\arabic{figure}}
\clearpage
\onecolumngrid
\appendix
\begin{center}
\large{\bf{Supplemental Materials for\\
Extraordinary $\pi$-Electron Superconductivity\\
Emerging from a Quantum Spin Liquid
}}
\end{center}
\section{Experimental methods}
The single crystals of $\kappa$-(BEDT-TTF)$_4$Hg$_{2.89}$Br$_8$ were electrochemically synthesized.
50~mg of Bu$_4$NHgBr$_3$, 10~mg of HgBr$_2$, and 10~mg of BEDT-TTF were placed in an H-shaped electrochemical cell with 15~mL of distilled 1,1,2-tetrachroloethene.
A galvanostatic current of 0.5~$\mu$A was applied and black block crystals were harvested as minor products after 6 months.
The typical sizes of the gained crystals are 300$\times$300$\times$50~$\mu$m$^3$.
The obtained crystals were identified by x-ray diffraction and the directions of the crystal axes were simultaneously determined.
In this report, we assume “1 mol” as “1 mol of dimer” for comparison with a series of typical BEDT-TTF-based organic compounds, such as $\kappa$-(BEDT-TTF)$_2$X salts.

The electrical resistance was measured by the conventional four-terminal method in a 17~T superconducting magnet with a $^4$He VTI system, a 20~T $^3$He/$^4$He dilution refrigerator, and a 55~T pulse magnet with a $^3$He cryostat.
The tunnel diode oscillator (TDO) measurements were performed by using a tunnel diode oscillator with the sample mounted in an 8-shaped coil.
The typical resonant frequency is $\sim$82~MHz, which gives the skin depth of the normal state of $\kappa$-HgBr as $>$1~mm due to the low conductivity\cite{9,10,11}.
Since the depth is sufficiently larger than the thickness of the sample (about 100~$\mu$m), the frequency change predominantly reflects the superconductivity.

The magnetic susceptibility of polycrystalline samples of 1.116~mg was measured by a SQUID (Quantum Design 1~T MPMS).
No obvious Curie-type paramagnetic component was detected.
The magnetic susceptibility was obtained by subtracting the core-diamagnetism from the measured susceptibility.
The magnetic torque was measured by a microcantilever technique\cite{37}.
The samples used in the measurements were prepared by slicing the as-grown crystals and the sliced fragments were attached to the micro-cantilevers with grease along each crystal axis.

The low-temperature heat capacity measurements were performed by using a high-resolution thermal relaxation calorimeter\cite{38} in a 15~T superconducting magnet with $^3$He refrigerator.
We used a single crystal of $\kappa$-HgBr that had a weight of 309.1~$\mu$g.
Addenda heat capacity was determined in a separate run performed before mounting the sample.
The absolute value of the sample heat capacity was obtained by subtracting the background values from the total heat capacity.

\newpage
\section{Definition of $H_{\rm f}$ in the resistance measurements}
\begin{figure}[hh]
\begin{center}
\includegraphics[width=0.7\linewidth,clip]{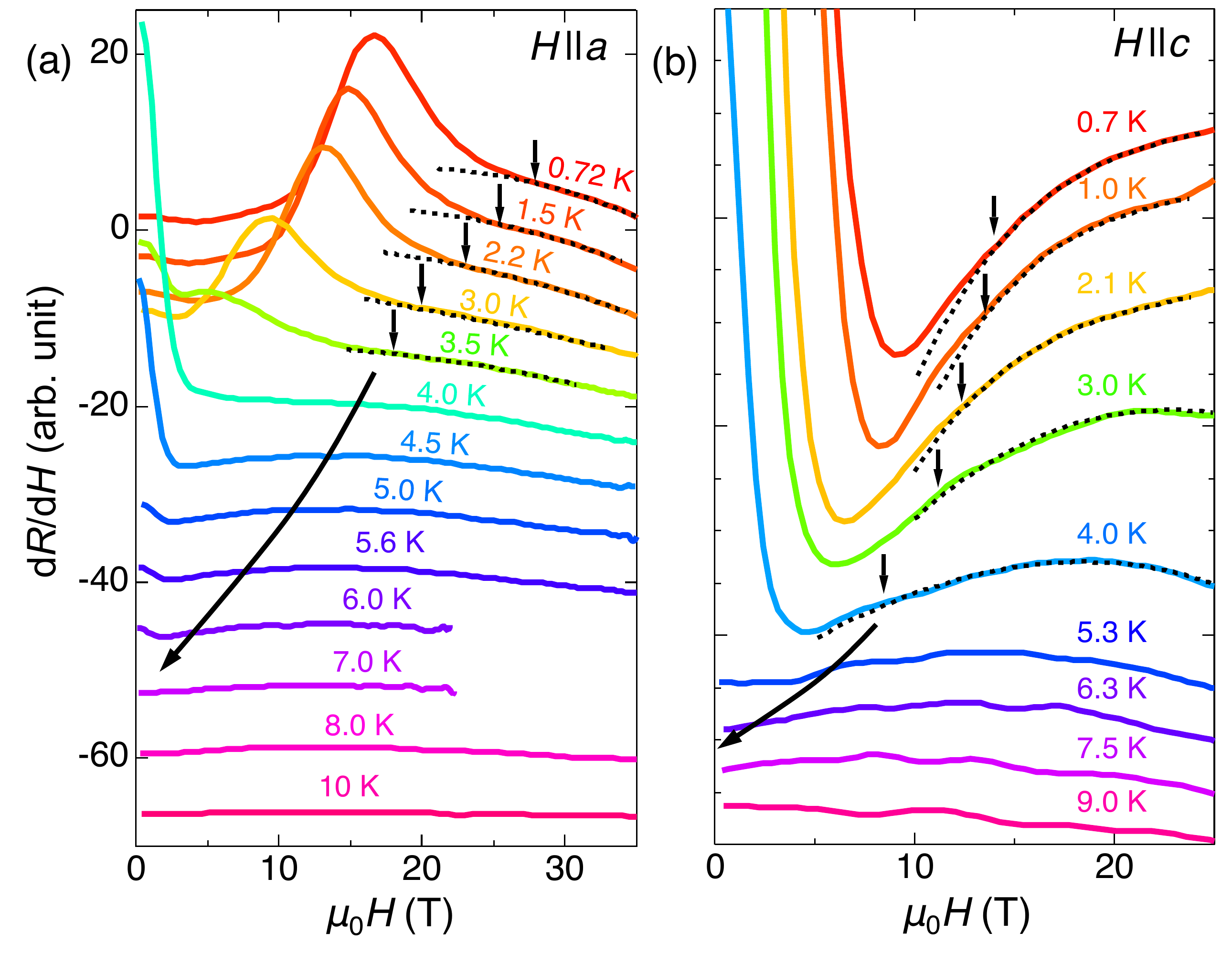}
\end{center}
\caption{
(a),(b) Field derivative of the out-of-plane magnetoresistance in parallel fields (a) and perpendicular fields (b) to the conducting plane.
The dashed lines are the binomial fits of the field-derivative magnetoresistance of the normal state.
The small arrows represent the onset fields to begin to deviate from the normal state.
The long arrows are guides for the eye to show the rough estimate of $H_{\rm f}$ because the determination is difficult at higher temperatures.
}
\label{figS1}
\end{figure}

In the present electrical transport measurements, the emergence of the fluctuating superconductivity is determined as the fields to begin to deviate from the field derivative of the magnetoresistance d$R$/d$H$ of the normal state, which means the onset, as indicated the black arrows in Fig.~\ref{figS1}(a),(b).
The field derivative of the magnetoresistance of the normal state is estimated by a simple binomial fit.
The broadened resistive transition makes the accurate determination difficult at higher temperatures.
Since the determined $H_{\rm f}$ is well consistent with the $H_{\rm f}$ given in the TDO measurements and the torque magnetometry (see Fig. 5), the definition seems to make sense for the present sample.

\section{Estimation of magnetic susceptibility}
\begin{figure}
\begin{center}
\includegraphics[width=0.9\linewidth,clip]{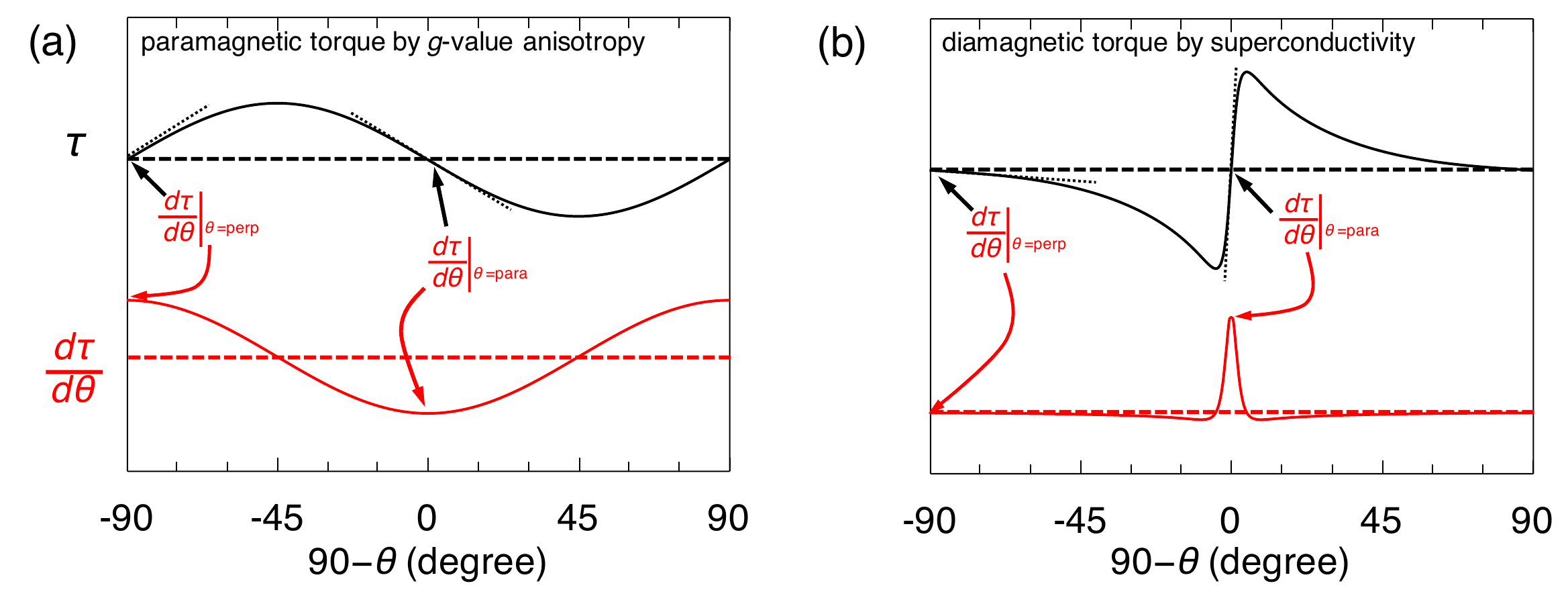}
\end{center}
\caption{
Angle dependences of (a) the paramagnetic torque which arises from the anisotropy of the $g$-factor and (b) the diamagnetic torque originating from the superconductivity (the black curves) with angle dependences of their angle derivatives (the red curves).
The dotted black lines indicate the slopes of the torque curves at parallel and perpendicular direction, which correspond to the amplitude of the angle derivatives.
}
\label{figS2}
\end{figure}

In Fig.~\ref{figS2}, we show the angle dependences of the paramagnetic torque produced by the $g$-value anisotropy and the diamagnetic torque originating from the superconductivity with the angle dependences of their angle derivatives.
The angle and the magnetic field dependences of the magnetic torque $\tau$ of the paramagnetic contribution is expressed by the next formula\cite{39,40},
\begin{equation}
\tau_{a{\rm -}c}(\theta,T,H)=-\frac{1}{2}\{\chi_{1}(T)-\chi_{2}(T)\}\mu_{0}H^{2}V{\rm sin}(2\theta-\theta_{0}),
\end{equation}
\begin{equation}
\tau_{a{\rm -}b}(\theta^{\prime},T,H) =-\frac{1}{2}\{\chi_{1}(T){\rm cos}^{2}\theta_{0}+\chi_{2}(T){\rm sin}^{2}\theta_{0}-\chi_{3}(T)\}\mu_{0}H^{2}V{\rm sin}(2\theta^{\prime}),
\end{equation}
where $\chi_{\rm i}$, $\mu_{0}$, $V$, and $\theta_{0}$ represent the magnetic susceptibility along each principal axis, the magnetic permeability, the sample volume, and the angle from the long axis of BEDT-TTF molecules and $a$-axis, respectively.
This expression succeeded to reproduce angle dependence of paramagnetic components in the other QSL materials\cite{39,40,14}.
Moreover, since temperature dependence of the anisotropy of the susceptibility of $\kappa$-HgBr is almost negligible\cite{9,41}, the angle derivative of the torque at $a$-axis is given as
\begin{equation}
d\tau_{a{\rm -}c}(T)/d\theta|_{\theta=a}=\{\chi_{1}(T)-\chi_{2}(T)\}\mu_{0}H^{2}V{\rm cos}(\theta_0)\propto\mu_{0}\chi(T)H^{2}|_{H\parallel a},
\end{equation}
\begin{equation}
d\tau_{a{\rm -}b}(T)/d\theta^{\prime}|_{\theta^{\prime}=b}=\{\chi_{1}(T){\rm cos}^{2}\theta_{0}+\chi_{2}(T){\rm sin}^{2}\theta_{0}-\chi_{3}(T)\}\mu_{0}H^{2}V\propto\mu_{0}\chi(T)H^{2}|_{H\parallel b}.
\end{equation}
Indeed, the temperature dependence of the magnetic susceptibility obtained by the formulae qualitatively reproduces that derived by the SQUID magnetometry as shown in Fig. 3(b).
Thus, we evaluate the absolute value of the susceptibility of the paramagnetic component from the normalization to $\chi$ determined by the SQUID magnetometry to avoid the ambiguity in the weight of the tiny single crystals used in the torque measurements.
Although the magnitude of applied fields is different, the temperature dependences of the overlapped region of the measurements are qualitatively same, meaning that the magnetic susceptibility in the normal state does not have magnetic field dependence as in the case of the Pauli paramagnetism.
After subtracting the paramagnetic components from the total torque curves, we evaluate the diamagnetic susceptibility of the superconductivity $\chi_{\rm sc}$($T$) in parallel fields by the angle derivative of the torque because it can be described by the following formula\cite{42} using the diamagnetic susceptibility $\chi_{\rm sc}$ and the diamagnetism $M_{\rm sc}$,
\begin{equation}
d\tau/d\theta|_{\rm para}=-\mu_0H_{\rm para}d(M\times H)/dH_{\rm perp}|_{\rm para}=-\mu_0H_{\rm para}d(H_{\rm para}M_{\rm perp}-M_{\rm para}H_{\rm perp})/dH_{\rm perp}|_{\rm para}.
\end{equation}
Here, we use these relations, $H_{\rm para}$=$H$sin$\theta$ and $H_{\rm perp}$=$H$cos$\theta$. When fields are applied along the nearly in-plane direction ($\theta$$\sim$90 deg.), $M_{\rm para}$$H_{\rm perp}$ is negligible.
Thus,
\begin{equation}
d\tau/d\theta|_{\rm para}\approx-\mu_0dM_{\rm perp}/dH_{\rm perp}|_{\rm para}H_{\rm para}^2\propto\mu_0\chi_{\rm sc}(T)H^2.
\end{equation}
The “perp”, ”para”, and ”sc” in the subscripts and superscripts are the abbreviations of ”perpendicular”, ”parallel”, which mean the directions to the conducting layers, and “superconductivity”, respectively.
Imitating this method, $\chi_{\rm sc}$($T$) in perpendicular fields can be also determined by the angle derivative of the torque although the sign and the coefficient are different.
Also, it should be noticed that the diamagnetism cannot be scaled with the obtained Pauli paramagnetism because the coefficients of the absolute value are different.
In the normal state, since the diamagnetic susceptibility does not exist, the calculated values of $\chi_{\rm sc}$($T$) must be zero, which means that those are in the noise floor as described by the shaded areas in the semi-logarithmic plots in Figs. 3(c) and 3(d).

\section{Irreversibility field of the superconductivity}
\begin{figure}[hh]
\begin{center}
\includegraphics[width=0.8\linewidth,clip]{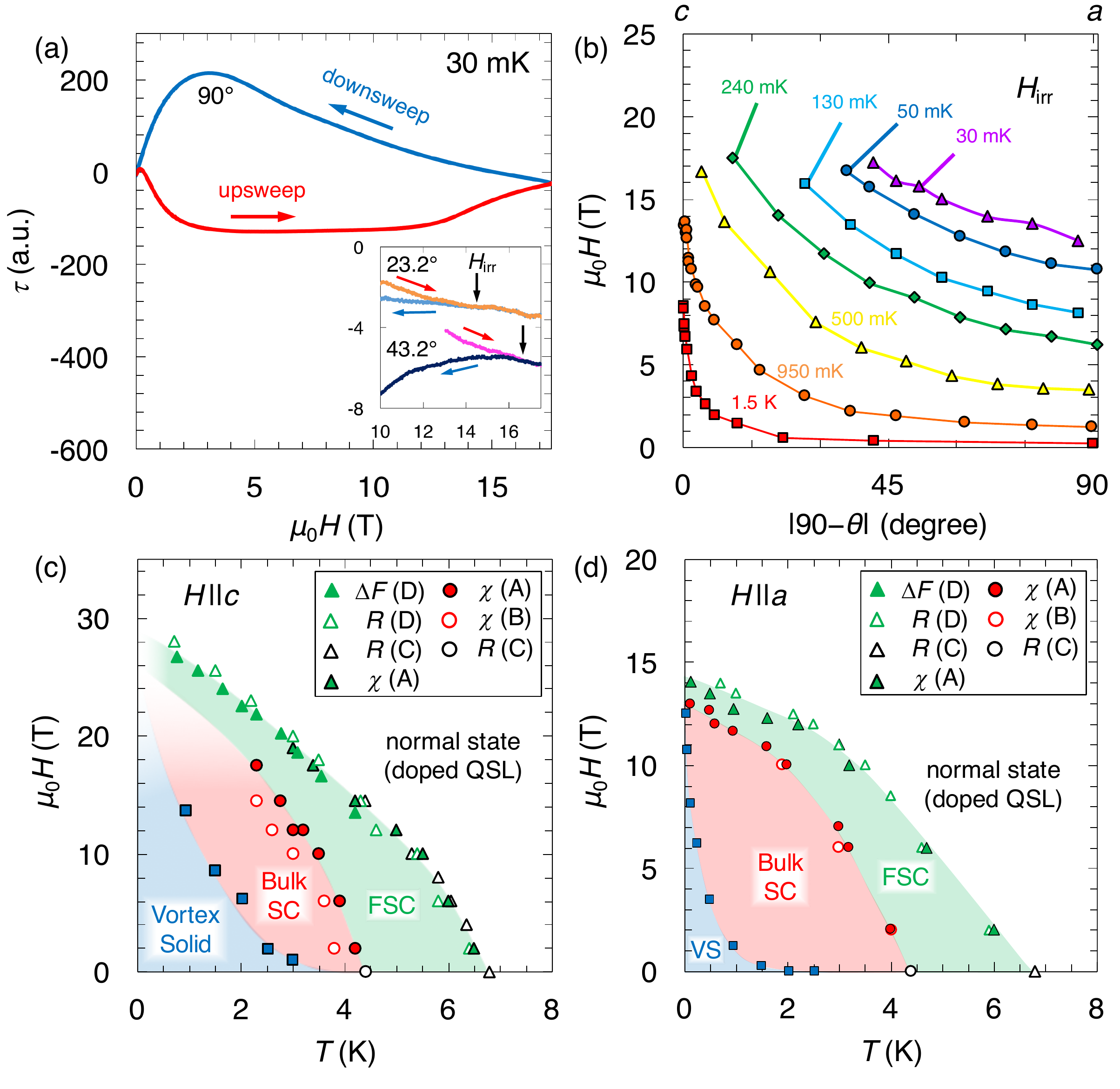}
\end{center}
\caption{
(a) Magnetic field dependences of the torque in upsweep and downsweep processes of fields at 90 deg. ($c$-axis direction) at 30~mK.
The inset shows angle variations of the field dependences.
The black arrows indicate $H_{\rm irr}$ at each angle.
(b) Angle dependences of $H_{\rm irr}$ at each temperature.
(c),(d) The updated superconducting phase diagram of Fig. 5(a) and 5(b) for (c) parallel and (d) perpendicular fields.
The blue shaded areas represent vortex-solid (VS) state.
The blue boxes denote $H_{\rm irr}$ at each temperature.
}
\label{figS3}
\end{figure}

The magnetic torque curves at 30~mK in Fig.~\ref{figS3}(a) show large hysteresis originating from the vortex pinning in superconducting layers.
The field where the hysteresis disappears is defined as the irreversibility field $H_{\rm irr}$, which is predominantly influenced by the surface pinning.
The vortex melting field $H_{\rm m}$ is expected to be slightly lower than or almost equal to $H_{\rm irr}$\cite{43}.
Also, $H_{\rm m}$ should be lower than the critical field of the bulk superconductivity $H_{\rm c2}$.
We present the angle dependence of $H_{\rm irr}$ at each temperature in Fig.~\ref{figS3}(b).
Although the behavior is very similar to angle dependence of $H_{\rm c2}$ for anisotropic superconductors, the anisotropy is not explained by the 2D model\cite{44} or the anisotropic 3D model\cite{45} which are often applied to describe the anisotropy of $H_{\rm c2}$ because of the presence of the effect of the surface pinning.
Figures~\ref{figS3}(c) and \ref{figS3}(d) are the updated superconducting phase diagram of Fig. 5a and 5b including these results.

\section{Estimation of electronic heat capacity}
\begin{figure}[hh]
\begin{center}
\includegraphics[width=0.9\linewidth,clip]{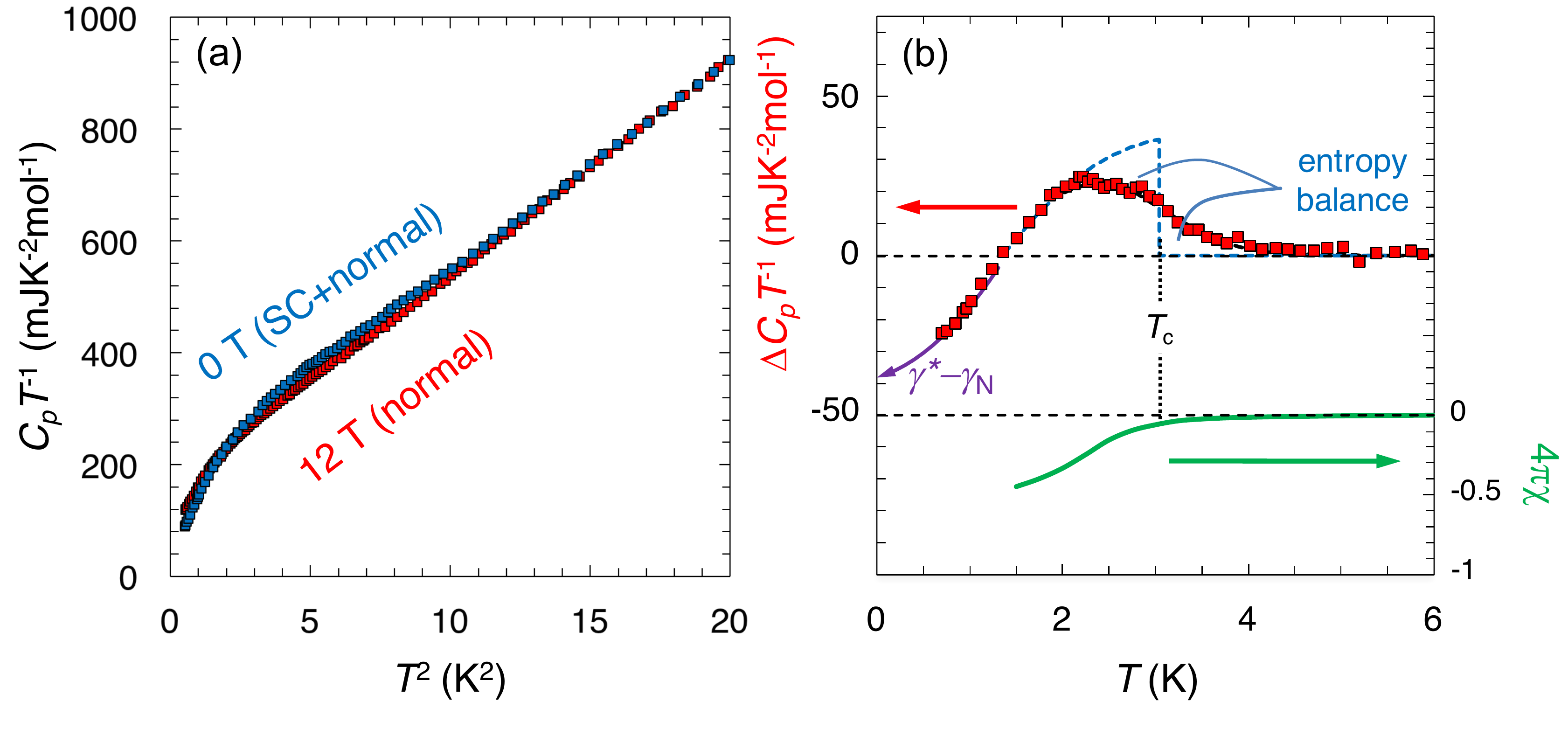}
\end{center}
\caption{
(a) Temperature dependence of the heat capacity up to 20~K$^2$ at 0~T and 12~T.
(b) The thermodynamic anomaly related to the superconducting transition and Meissner volume as a function of temperature\cite{11}.
}
\label{figS4}
\end{figure}

The obtained heat capacities in the present study are well consistent with the data of the previous work\cite{22}.
Figure~\ref{figS4}(a) shows the temperature dependence of the heat capacity as a $C_pT^{-1}$ vs $T^2$ plot.
In the case of conventional metals, temperature dependence of heat capacity is described as $C_p$=$\gamma_{\rm N}$$T$+$\beta$$T^3$, where the first term $\gamma_{\rm N}$$T$ is electronic heat capacity and the second $\beta$$T^3$ represents Debye-type lattice heat capacity.
Thus, the temperature dependence should show a linear relation in this plot and the intercept and slope gives $\gamma_{\rm N}$ and $\beta$, respectively.
However, the heat capacity of the normal state of $\kappa$-HgBr displayed in Fig. 4a and ~\ref{figS4}(a) does not show such linear behavior due to the presence of some additional term.
The previous work\cite{22} suggests that the excess heat capacity is attributed to dimensional crossover of phonon of the Hg chain in the anion layers because the acoustic phonon of the heavy atoms Hg can have strong anisotropy even at low temperatures.
Indeed, they succeed to reproduce the temperature dependence by using the Tarasov model\cite{46}, which describes such dimensional crossover of phonon.
However, we should note that there is another possibility that the QSL state has a characteristic low-energy excitation different from the conventional electronic heat capacity $\gamma_{\rm N}$$T$ as some works\cite{47,48} suggest.
We thus use the simple extrapolation of the polynomial formula up to 4th order (the dotted line in Fig. 4a) to estimate the electronic heat capacity coefficients $\gamma_{\rm N}$ and $\gamma^{\ast}$.
In either case, we can evaluate the electronic heat capacity related to the superconducting state from the field-dependent term since the normal state of the salt should have no magnetic field dependence.

Next, we focus on the discussion of the superconductivity.
Figures 4b and \ref{figS4}(b) shows the temperature dependence of the superconducting electronic heat capacity as the difference between the 0~T and 12~T data.
As we can see, the transition is much broadened by the strong quantum fluctuation.
Typically, small broadening by thermal fluctuation around $T_{\rm c}$ can be corrected by introducing a Gaussian.
However, it is difficult for the present case to correct the transition behavior because the superconductivity is no longer treated in the framework of the mean-field.
The thermodynamic anomaly cannot be reproduced by any existing models, such as the typical BCS behavior and the $\alpha$-model\cite{25} which is often applied to describe the heat capacity of unconventional superconductors.
Thus, we simply use a binomial fit (blue dashed line in Fig.~\ref{figS4}(b)) to correct the fluctuation effect with taking the entropy balance into consideration.
Here, $T_{\rm c}$ is determined by the entropy balance of the areas indicated in Fig. 4b.
The determined $T_{\rm c}$ from the heat capacity seems to be lower than the $T_{\rm c}$ measured by the torque magnetometry because of the difference of the definition.
The temperature 4.2~K, that is Tc determined by the torque measurements, is the onset in the heat capacity.
These also indicate the difficulty of the analysis of the broadened transition by the typical method.


\end{document}